\documentstyle[preprint,eqsecnum,aps,epsf]{revtex}	

\newif\iftightenlines\tightenlinesfalse
\tightenlines\tightenlinestrue

\begin{document}
%
\def\pT{p_T^{\phantom{7}}}
\def\MW{M_W^{\phantom{7}}}
\def\ET{E_T^{\phantom{7}}}
\def\bh{\bar h}
\def\lm{\,{\rm lm}}
\def\lo{\lambda_1}
\def\lt{\lambda_2}
\def\pslt{p\llap/_T}
\def\eslt{E\llap/_T}
\def\to{\rightarrow}
\def\Re{{\cal R \mskip-4mu \lower.1ex \hbox{\it e}}\,}
\def\Im{{\cal I \mskip-5mu \lower.1ex \hbox{\it m}}\,}
\def\SU{SU(2)$\times$U(1)$_Y$}
\def\te{\tilde e}
\def\tl{\tilde l}
\def\tb{\tilde b}
\def\tevst{TeV$^*\ $}
\def\ttau{\tilde \tau}
\def\tg{\tilde g}
\def\tga{\tilde \gamma}
\def\tnu{\tilde\nu}
\def\tell{\tilde\ell}
\def\tq{\tilde q}
\def\tst{\tilde t}
\def\tw{\widetilde W}
\def\tz{\widetilde Z}
\def\cmsec{{\rm cm^{-2}s^{-1}}}
\def\sgn{\mathop{\rm sgn}}
\hyphenation{mssm}
\def\ds{\displaystyle}
\def\ts{${\strut\atop\strut}$}
%
\preprint{\vbox{\baselineskip=14pt%
   \rightline{FSU-HEP-950301}\break
   \rightline{UR-1411}\break
   \rightline{ER-40685-858}
   \rightline{UH-511-826-95}
}}
\title{SUPERSYMMETRY REACH OF AN UPGRADED TEVATRON COLLIDER}
\author{Howard Baer$^1$, Chih-hao Chen$^1$, Chung Kao$^2$ and Xerxes Tata$^3$}
\address{
$^1$Department of Physics,
Florida State University,
Tallahassee, FL 32306, U.S.A.}
\address{
$^2$ Department of Physics and Astronomy, University of Rochester, Rochester,
NY 14627, U.S.A.}

\address{
$^3$Department of Physics and Astronomy,
University of Hawaii,
Honolulu, HI 96822, U.S.A}
\date{\today}
\maketitle
\begin{abstract}
We examine the capability of a $\sqrt{s}=2$ TeV Tevatron $p\bar p$ collider
to discover supersymmetry, given a luminosity upgrade to amass $25\ fb^{-1}$
of data. We compare with the corresponding reach of the Tevatron
Main Injector ($1\ fb^{-1}$ of data).
Working within the framework of minimal supergravity with
gauge coupling unification and radiative electroweak symmetry breaking,
we first calculate the regions of parameter space accessible via the clean
trilepton signal from $\tw_1\tz_2\to 3\ell +\eslt$ production, with detailed
event generation of both signal and major physics backgrounds.
The trilepton signal can allow
equivalent gluino masses of up to $m_{\tg}\sim 600-700$ GeV
to be probed if $m_0$ is
small. If $m_0$ is large, then $m_{\tg}\sim 500$ GeV can be probed for
$\mu <0$; however, for $\mu >0$ and large values of $m_0$, the rate for
$\tz_2\to\tz_1\ell\bar{\ell}$ is suppressed by
interference effects, and there is {\it no} reach in this channel.
We also examine regions where the signal from
$\tw_1\overline{\tw_1}\to \ell\bar{\ell}+\eslt$ is detectable.
Although this signal
is background limited, it is observable in some regions where
the clean trilepton signal is too small.
Finally, the signal $\tw_1\tz_2\to jets+\ell\bar{\ell} +\eslt$
can confirm the clean trilepton signal in a substantial subset
of the parameter space where the trilepton signal can be seen.
We note that
although the clean trilepton signal may allow Tevatron experiments to identify
signals in regions of parameter space beyond the reach of LEP II, the dilepton
channels generally probe much the same region as LEP II.

\end{abstract}

\medskip



\section{Introduction}

The search for weak scale supersymmetric particles is a high priority item for
colliding beam experiments\cite{MSSM,DPF}. Supersymmetric models with
sparticles of mass $\sim 100-1000$ GeV are known to
stabilize the Higgs boson mass against quantum corrections which tend to
escalate $m_H$ to some ultra-high mass scale when the Standard Model (SM) is
embedded within a larger framework.
Furthermore, weak scale
supersymmetry (SUSY) allows for a simple unification of gauge coupling
constants at a unification scale $M_U\sim 2\times 10^{16}$ GeV\cite{UNIF}.
So far, no direct evidence for supersymmetry has been
found either at LEP\cite{LEP}
(where sparticle mass limits of $\sim {M_Z\over 2}$ have been obtained),
or at Tevatron experiments
(which have excluded gluinos and squarks lighter than about
150~GeV\cite{CDFDZ}).

In the future, LEP II is expected to explore sparticle masses up to
$m_{H_\ell},\ m_{\tw_1},\ m_{\tst_1},\ m_{\tell}\sim 90$ GeV\cite{DIONISI,DPF}.
The Fermilab
Tevatron experiments will be able to explore further ranges of $m_{\tg}$ and
$m_{\tq}$ by searching for multi-jet$+\eslt$ events and multi-lepton $+$
multi-jet $+\eslt$ events from gluino and squark cascade decays\cite{BKT}.
Even if gluinos and most of the squarks are beyond the reach of
Tevatron collider experiments, it may still be possible to find
signals for the top squark\cite{BST}, which is expected\cite{ER} to have a
mass lighter than, and sometimes much lighter than,
the other five flavors of squarks.

There has recently been much interest in the clean trilepton signal from
$\tw_1\tz_2\to 3\ell$ at the Tevatron
collider\cite{EARLY,ANTRI,BARB,NEW,LOPEZ,KAMON,KANE,GUNION,CDF,DZERO}.
Many studies within the
minimal supergravity (SUGRA) framework, find that
$m_{\tw_1}\simeq m_{\tz_2}\sim ({1\over 4}-{1\over 3}) m_{\tg}$;
it may thus
be that gluinos and squarks are beyond the reach of Tevatron experiments,
while charginos and neutralinos are produced with significant cross sections.
In addition, the $\tz_2$ (and sometimes also the $\tw_1$) frequently have
large branching fractions to leptons.
This is in part because slepton masses are
typically smaller than squark masses, so that decays via virtual sleptons
are enhanced. Also, in some regions of parameter space, sparticle
mass and mixing patterns and/or
interference effects can enhance (but also diminish) the leptonic
branching fraction of $\tz_2$,
even if $m_{\tell}\sim m_{\tq}$.
Combining production cross sections with branching ratios shows
that the clean $3\ell +\eslt$ signal can be substantial over large regions
of parameter space, whereas SM backgrounds for this topology are expected
to be tiny.

Already the CDF and D0 experiments at the Tevatron collider have collectively
accumulated an integrated
luminosity in excess of 0.1 $fb^{-1}$ and should amass a considerably larger
data sample by the end of the
current run, run IB. This
should allow an
exploration of $m_{\tg}\sim 200$ GeV ($m_{\tg}\sim 250$ GeV) for
$m_{\tq}>>m_{\tg}$ ($m_{\tq}\sim m_{\tg}$), via the multi-jet$+\eslt$
search\cite{DPF}.
In addition, experiments will
finally be sensitive to exploring some of parameter space via top squark
searches, and via clean trilepton searches. The Tevatron Main Injector (MI)
is expected to begin operation in 1999, at $\sqrt{s}=2$ TeV,
and should allow for integrated
luminosities of $\sim 1\ fb^{-1}$ to be accumulated per experiment.
Recently, there have been discussions concerning further luminosity
upgrades beyond the MI, via anti-proton recycling and storage. These
further luminosity upgrades, dubbed the \tevst (TeV-star) project,
could possibly accumulate $\sim 25fb^{-1}$ of integrated luminosity
per experiment. In this paper, we evaluate
the capability of the MI and \tevst to search for sparticles,
and compare their reach with that of LEP II and the recently approved LHC.

We work within the framework of the minimal SUGRA
model\cite{SUG} with radiative breaking of electroweak symmetry.
Several groups\cite{RGE} have studied the expectations for sparticle masses
within this rather restricted framework which is completely determined
by just four SUSY parameters renormalized at some ultra-high scale where the
physics, because of assumptions about the symmetries of the interactions, is
simple. These may be taken to be,
\begin{itemize}
\item $m_0$, a common scalar mass term,
\item $m_{1/2}$, a common gaugino mass term,
\item $A_0$, a common tri-linear coupling, and
\item $B_0$, a bilinear coupling.
\end{itemize}
The various gauge and Yukawa couplings and soft breaking terms
are then evolved from the unification scale down to the weak scale. The
evolution\cite{INOUE} can be traced via 26 renormalization group equations
(RGE's).
A remarkable consequence of this mechanism is that electroweak
symmetry is automatically broken when one of the Higgs mass squared terms
gets driven to a negative value. The correct symmetry breaking
pattern is then obtained for a large range of model parameters.
The bilinear coupling $B_0$ can be
eliminated in favor of $\tan\beta$ (ratio of Higgs field vev's), and the
magnitude (but not the sign) of the supersymmetric Higgs mass term $\mu$
can be solved for in terms of $M_Z$. Hence, the complete weak scale
sparticle spectrum and sparticle mixing angles
can be calculated in terms of the parameter set
\begin{eqnarray}
m_0,\ m_{1/2},\ A_0,\ \tan\beta ,
\end{eqnarray}
together with the sign of $\mu$ and the
top quark mass $m_t$.
An iterative solution of the RGE's,
using two-loop gauge coupling RGE's and the one-loop effective potential,
has been implemented as the subprogram ISASUGRA\cite{BCMPT},
and is a part of the ISAJET
package\cite{ISAJET}, which we use in our analysis.

In order to assess the best channels for SUSY discovery,
we first present total cross sections
for various sparticle pair-production mechanisms in Fig.~1, for $p\bar p$
collisions at $\sqrt{s}=2$ TeV, using CTEQ2L parton distribution
functions\cite{CTEQ}. To be specific, we show results for $\tan\beta =2$
and a large and negative $\mu$ value, typical in SUGRA models. For Fig. 1{\it
a},
we show total cross sections versus $m_{\tg}$ for $m_{\tq}=m_{\tg}$.
We see that for $m_{\tg}<300$ GeV, strong production of $\tg\tg$, $\tg\tq$
and $\tq\tq$ pairs dominates. However, as $m_{\tg}$ increases, the strong
production cross sections drop steeply, and are ultimately overtaken
by those for the production of lighter charginos and neutralinos via the
$p\bar p\to\tw_1\tz_2$ and $p\bar p\to\tw_1\overline{\tw_1}$ reactions. These
processes
dominate for $m_{\tg}>300$ GeV. Also shown is the total cross
section\cite{ASSOC}
for gluinos and squarks produced in association with charginos and
neutralinos (labelled assoc. prod.); this class of reactions is always
sub-dominant below strongly interacting pair production or chargino/neutralino
pair production. In Fig. 1{\it b}, we plot the same cross sections, except
that now $m_{\tq}=2m_{\tg}$. Again, strongly interacting sparticle pair
production (along with on-shell $W\to\tw_1\tz_1$ production)
is dominant for small $m_{\tg}$, but this time it becomes
sub-dominant for $m_{\tg}>200$ GeV.

The cross sections for a large, positive value of
$\mu$ are shown in Fig.~2. In this
case, $\tw_1$ and $\tz_{1,2}$ tend to be significantly lighter than for
negative values of $\mu$, so that their production cross
sections are even larger and dominate the strong production
cross sections for smaller gluino masses than in Fig.~1.
Notice that, for $\mu > 0$,
the non-observation of any SUSY signals
at LEP already indirectly excludes gluinos up to about 250~GeV.

The implications from Figs.~1 and 2 are clear:
for current values of integrated
luminosities, when the
Tevatron is exploring $m_{\tg}<200-300$ GeV, experiments should obtain
maximal reach by looking for signals from strongly interacting $\tg\tg$,
$\tg\tq$ and $\tq\tq$ production, such as multi-jets + $\eslt$ events, and
isolated leptons produced {\it in association} with jets from cascade decays
of gluinos and squarks\cite{BKT}.
To probe values of $m_{\tg}\agt 300$ GeV, which should be possible
with larger data samples, Tevatron experiments
should focus on signatures from $\tw_1\tz_2$ and $\tw_1\overline{\tw_1}$
production\cite{FN1}. These reactions will also determine the ultimate
reach of experiments at the Tevatron, provided that the decay patterns
of the charginos and neutralinos lead to final states that are observable
above SM backgrounds.
There are only a few promising experimental signatures from
$\tw_1\tz_2$ and $\tw_1\overline{\tw_1}$ production. The previously mentioned
clean trilepton signal from $\tw_1\tz_2\to 3\ell +\eslt$ occurs at an
observable
rate in large regions of the SUGRA parameter space and has only tiny
{\it physics} backgrounds. There are also
clean isolated dilepton signatures from
$\tw_1\overline{\tw_1}\to\ell\bar{\ell '}+\eslt$ production,
although this suffers from large SM background due to $WW$
production.
Furthermore, there is an isolated dilepton $+$jets signature from
$\tw_1\tz_2\to q\bar{q'}\tz_1 + \ell\bar{\ell}\tz_1$, which should be large
in much the same region where the $3\ell$ signature is visible, but which
also suffers from large backgrounds. These dilepton signals could serve to
confirm a signal in the relatively clean trilepton channel\cite{FN2}.
Finally, single isolated lepton $+$ jets
and multi-jet $+\eslt$ signatures from these reactions, and also
from $\tw_1\tz_1$ production,
should be buried
below backgrounds from direct $W$ production and QCD multi-jet production,
respectively.

To outline the rest of this paper, in Sec. II we show the results of
detailed and extensive calculations of signal and background for the
$3\ell +\eslt$
events, and plot the reach of the Tevatron MI and also \tevst in SUGRA
parameter space. In Sec. III, we perform a similar analysis for
$\tw_1\overline{\tw_1}\to\ell\bar{\ell} +\eslt$ events, and in Sec. IV, we
consider
$\tw_1\tz_2\to jets+ \ell\bar{\ell}+\eslt$ events.
In Sec. V, we combine the results of the
previous three sections with studies of signals from gluino and squark cascade
decays, to compare the SUSY search capabilities of the Main Injector and
\tevst upgrades with one another, and also with LEP II and the LHC.

\section{Reach via the clean trilepton signal}

To calculate $3\ell +\eslt$ signal levels, we use the ISAJET 7.13
event generator program\cite{ISAJET}. We generate all possible
subprocesses for chargino and neutralino pair production in our
calculations, even though $q\bar q\to \tw_1\tz_2$ is the dominant
component. The various chargino and neutralino
decay branching ratios are calculated within ISAJET, to lowest order.
ISAJET neglects initial/final state spin correlations, but these are
not expected to be of great significance at hadron colliders. ISAJET also
neglects decay matrix elements in the event generation
(but not the branching fraction calculation); these are mainly expected to
be significant when virtual states in 3-body decays are getting close
to being on mass shell. ISAJET is also used to calculate the dominant
physics backgrounds, which are expected (after cuts) to be $t\bar t$
production, and $WZ$ production.

We model experimental conditions using a toy
calorimeter with segmentation $\Delta\eta \times \Delta\phi = 0.1 \times
0.09$ and extending to $|\eta| = 4$. We assume an energy resolution of
$\frac{0.7}{\sqrt{E}}$ ($\frac{0.15}{\sqrt{E}}$) for the hadronic
(electromagnetic) calorimeter. Jets are defined to be hadron clusters
with $E_T > 15$~GeV in a cone with
$\Delta R=\sqrt{\Delta\eta^2+\Delta\phi^2}=0.7$. Leptons with $p_T > 8$~GeV
and within $|\eta_{\ell}| < 3$ are considered to be isolated if the hadronic
scalar $E_T$ in a cone with $\Delta R = 0.4$ about the lepton is smaller
than $\frac{E_T(\ell)}{4}$. In our analyses, we neglect multiple
scattering effects, an explicit detector simulation, and fake backgrounds
from $\gamma$ and jet mis-identification.
For this reason, our results should be regarded as optimistic.

To extract signal from background, various sets of cuts have been
proposed\cite{NEW,KAMON,CDF,DZERO}. In this paper, since we are interested in
the maximal reach of a high luminosity Tevatron collider, we take
somewhat harder cuts than these previous studies. In Table I, we list
the cross section after successive cuts,
for two SUGRA cases: ($m_0,m_{1/2},A_0,\tan\beta ,sign(\mu ),m_t$)=
(200,165,0,2,-1,170) and (200,200,0,2,+1,170), where all mass parameters
are in GeV. We also list backgrounds from $t\bar t$ production for
$m_t=170$ GeV, and from $WZ$ production.

The cuts we implement are the following.
\begin{itemize}
\item We require 3 {\it isolated} leptons in each event, with $p_T(\ell_1)>20$
GeV, $p_T(\ell_2)>15$ GeV, and $p_T(\ell_3)>10$ GeV. In addition, each lepton
is central ($|\eta |<2.5$). This cut is labelled ($3\ell$). We see that already
the background from $t\bar t$ production falls below the $1\ fb$ level,
leaving a large $WZ$ background.
\item We require $\eslt >25$ GeV (labelled ($\eslt$)). This cut removes
backgrounds (not listed) from SM processes such as Drell-Yan dilepton
production, where an extra accompanying jet fakes a lepton.
\item To reduce background from $WZ$ production, we require that the
invariant mass of any opposite-sign, same flavor dilepton pair not reconstruct
the $Z$ mass, {\it i.e.} we require that
$|m(\ell\bar{\ell})-M_Z|\geq 10$~GeV. This cut
reduces $WZ$ background to below the $1\ fb$ level, and is labelled ($M_Z$).
\item We finally require the events to be {\it clean}, {\it i.e.} no jets
(as defined above) should be present. This is labelled as ($0j$).
\end{itemize}
At this point, the surviving total background rates are $\sim 0.2\ fb$,
leaving $\sim 5$ background events per $25\ fb^{-1}$ of integrated luminosity.
The background mainly comes from $WZ$ events, where $Z\to\tau\bar{\tau}$,
with subsequent $\tau$ leptonic decay. The $5\sigma$ level for discovery of a
signal corresponds to $0.45\ fb$, or $11$ events for the same luminosity.
The two signal cases listed in the table are well above this limit.

Our next task is to delineate the regions of SUGRA parameter space
where a clean trilepton signal is observable, both for the Tevatron MI,
as well as for TeV$^*$, after our simulation of signal events,
with the effect of cuts included. We show in Fig. 3{\it a} the regions of
interest in the $m_0\ vs.\ m_{1/2}$ plane, taking $A_0=0$, $\tan\beta =2$,
and $\mu <0$. Varying $A_0$ mainly affects third generation sparticle masses,
and so has little effect on our results, unless $\ttau_1$, $\tb_1$ or
$\tst_1$ are driven to such low values that new two-body decay modes
of charginos and neutralinos open up. We take $m_t=170$ GeV, consistent
with the values recently obtained by the CDF and D0 experiments\cite{CDFTOP}.
The regions of the plane shaded by a brick wall are excluded on theoretical
grounds, either because of lack of appropriate electroweak symmetry breaking,
or because the LSP becomes either the sneutrino (excluded in Ref.
\cite{BDT}, under the assumption that the LSP constitutes the galactic dark
matter),
or is charged (usually the $\tw_1$, $\te_R$ or $\ttau_1$). The
diagonally shaded region is excluded by various experimental
constraints, including the LEP limits\cite{LEP} of $m_{H_\ell}\agt 60$ GeV,
$m_{\tw_1}>47$ GeV, and $m_{\tnu}>43$ GeV, and the CDF and D0 limits
on $m_{\tg}$ and $m_{\tq}$ from multi-jet$+\eslt$ searches\cite{CDFDZ}.

The sampled points in SUGRA space which yield an observable signal at the
Tevatron MI, assuming $1\ fb^{-1}$ of integrated luminosity, are shown
with black squares. At the MI, where the probability
of getting more than one
background event is smaller than 2\%,
we have required a minimum of five signal events to
claim discovery---the Poisson probability for the {\it physics} background
to fluctuate to this level is $2\times 10^{-6}$.

To obtain our signal cross section, we generate all possible
chargino/neutralino production mechanisms. For each point, we
typically generate events until
either 25 signal events are produced, or a maximum of 50K events are generated,
or the signal cross section falls well below the $3\sigma$ level.

To facilitate the translation of the
points in the SUGRA $m_0\ vs.\ m_{1/2}$  plane
into sparticle masses, we show in Fig. 3{\it b} various sparticle mass
contours for $\tg$, $\tw_1$ and $\tell_R$. By comparing Fig. 3{\it a}
with 3{\it b}, we see that the maximal reach in $m_{\tg}$ at the MI is
achieved at $m_{1/2}\sim 160$ GeV, for which $m_{\tg}\sim 450$ GeV,
and $m_{\tw_1}\sim 150$ GeV. In this region of relatively low values of
$m_0$, sleptons are much lighter than squarks, so leptonic branching
ratios of $\tz_2$ are enhanced. For larger values of $m_{1/2}$, the
$\tz_2 -\tz_1$ mass difference becomes sufficiently large that the
two body decay mode $\tz_2\to\tz_1 H_{\ell}$ opens up, dominating
the $\tz_2$ branching fractions, and spoiling the signal. The onset of this
``spoiler mode'' is denoted by the correspondingly labelled
horizontal contour. If $m_0$ is sufficiently small, then sleptons and
sneutrinos become light enough that $\tz_2\to\tell_L\ell$,
$\tz_2\to\tell_R\ell$ or $\tz_2\to\tnu_L\nu$
are accessible, and can dominate $\tz_2$ decay modes. Diagonal
contours denoting the regions where these modes are
accessible are also labelled. We see that as $m_0$ increases, squarks
and sleptons become increasingly heavy. Even so, for large $m_0$, the Tevatron
MI still has a reach via clean trileptons out to $m_{1/2}\sim 60-80$ GeV,
corresponding to $m_{\tg}\sim 215-260$ GeV.

The SUGRA points accessible to \tevst are denoted by squares with x's
($10\sigma$ level), and open squares ($5\sigma$ level). Sampled points
with signal between 3-5$\sigma$ are labelled with triangles, and
points with less than a $3\sigma$ signal are denoted by x's.
Note that with the $5\sigma$ criterion of observability, the signal to
background ratio exceeds two. For small
$m_0$, we see that \tevst has an extraordinary reach out to
$m_{1/2}\sim 280$ GeV, corresponding to $m_{\tg}\sim 700$~GeV! In this
region, $\tz_2$, which is predominantly $SU(2)$-gaugino, dominantly
decays to $\nu\tnu_L$ and $\ell\tell_L$, even though $\tz_2\to\tz_1 H_{\ell}$
is accessible. As $m_0$ increases, the $\tz_2\to\ell\tell_L$ decay mode
closes, so $\tz_2\to\nu\tnu_L$ is nearly 100\%, and is invisible. This is
the cause for the region of non-observability that extends to low
$m_{1/2}$, to the left of the $\tz_2\to\nu\tnu_L$ contour. As $m_0$
increases further, the $\tz_2\to\nu\tnu_L$ decay is no longer accessible
and $\tz_2$ decays to $\ell\tell_R$,
although this mode now competes with the Higgs spoiler, which is typically
larger. For even larger values of $m_0$, all two-body decays to sleptons
are closed. In this region, \tevst can see (almost) up to the Higgs
spoiler over a large range of parameter space, corresponding to
$m_{1/2}\sim 170$~GeV, or $m_{\tg}\sim 500$~GeV, thus considerably expanding
upon the reach of the Tevatron MI.

We also see in Fig. 3{\it b} the region below the dashed contours which is
the approximate reach
of LEP II via Higgs, slepton pair or chargino pair searches.
We see that the MI and \tevst can probe significant regions beyond the LEP II
reach for charginos or sleptons, especially in the small $m_0$ area. However,
if MI or \tevst do see a signal, then LEP II probably should
have discovered the lightest SUSY Higgs boson.

In Fig.~4, we show the same results in the $m_0\ vs.\ m_{1/2}$ plane,
except now we choose $\mu  >0$. This mainly alters the
masses and mixing angles of the charginos and neutralinos from those in
Fig.~3.
In this case, we see that the Tevatron MI will have a maximal reach to
$m_{1/2}\sim 230$ GeV, corresponding to $m_{\tg}\sim 600$ GeV,
but only in the region where $m_0\alt 150$ GeV. The \tevst option
can again greatly expand this region, now probing up to $m_{1/2}\sim 280$
GeV ($m_{\tg}\sim 740$ GeV) in the favourable case when
$\tz_2$ decays to real sleptons are allowed.
If these decays are not allowed, then the region of detectability is limited
by the opening up of the $\tz_2\to Z\tz_1$ spoiler mode for the lower range
of $m_0$. Experiments at \tevst can probe
$m_0$ out to $m_0\sim 300-350$ GeV, corresponding
to $m_{\tell_R}\alt 400$ GeV. There is a major difference between the
positive (Fig. 4) and negative (Fig. 3) $\mu$ cases:
for larger values of $m_0$, there is {\it no reach at all
in $m_{1/2}$ beyond current LEP limits}, for $\mu >0$.
This lack of reach can be traced
directly to the $\tz_2\to\ell\bar{\ell}\tz_1$ branching fraction, which
can drop by two orders of magnitude when the sign of $\mu$ is flipped
to be positive, in the
large $m_0$ region\cite{FN}. The major decrease in $\tz_2$ branching fraction,
noted previously in Ref. \cite{NEW}, is mainly due to interference effects
amongst the $\tell_L$ and $Z$ mediated three body decay amplitudes. This
``hole'' lies in the region favored by the simplest
$SU(5)$ supergravity GUT model,
if we take proton decay and dark matter constraints\cite{AN} literally.
Such a scenario
can thus be well-tested at \tevst (via trileptons)
for $\mu <0$, but not for $\mu >0$.
Finally, we remark that for very large values of $m_0$ (and correspondingly
large slepton and squark masses) the amplitudes for
sfermion mediated decays of $\tz_2$ become very small, and the leptonic
branching fraction becomes the same as that for the $Z$ boson. The on-set
of this effect can be seen in Fig.~4 where we see that
for $m_0 \agt 800$~GeV, the crosses give way
to triangles and squares indicating the expected
reduction of the interference effects.

In Fig.~5, we show again the $m_0\ vs.\ m_{1/2}$ plane, but this time for
$\tan\beta =10$. We only show results for $\mu <0$ for brevity.
For positive values of $\mu$ and large values of $m_0$ very
similar results are obtained. This is because for large values of $\tan\beta$
the leptonic branching fraction
of $\tz_2$ is roughly symmetric about $\mu=0$ when its three body decays
dominate\cite{NEW} and the kinematics is roughly similar.
For small values of $m_0$ the results for positive and negative values
of $\mu$ are somewhat different. For $\mu < 0$ the neutralino mainly
decays invisibly via the $\tnu\nu$ mode whereas for positive values of
$\mu$ the $\tell_R\ell$ mode frequently dominates.

{}From Fig.~5, we see that
there are very few
points accessible to Tevatron MI. For large values of $m_0$,
this is mainly due to a large
enhancement in the hadronic $\tz_2$ decays, at the expense of leptonic
modes. In the small $m_0$ region, where dominantly two body decays take
place, most of the branching fraction is taken up by $\tz_2\to\tnu\nu$
decays. Although the branching fractions to the visible leptonic decays
are a few percent, it should be kept in mind that $m_{\tw_1}-m_{\tnu}$ is
rather small, so that the daughter lepton from chargino decay
tends to be soft resulting in a
reduced detection efficiency for these events. The corresponding region
may well be larger for positive values of $\mu$ and small $m_0$ where
the branching fraction for the leptonic $\tz_2$ decays are almost an order
of magnitude larger.
As in Fig.~3 and Fig.~4,
the \tevst collider option has a much greater reach throughout
parameter space than the Tevatron MI, and can explore up to $m_{\tg}\sim 700$
GeV, but only for a very narrow range of
$m_0\sim 160$ GeV. For $\mu < 0$ and
larger values of $m_0\sim 300-800$ GeV,
\tevst can probe to $m_{\tg}\sim 500$ GeV but the $5\sigma$ reach
frequently does not extend to where the spoiler decays become accessible.
Finally, for $m_0\sim 180-300$ GeV, there is again a ``hole'' of
non-observability extending all the way down to the current LEP constraint, due
again to the $\tz_2\to\ell\bar{\ell}\tz_1$ branching fraction suppression
(which also occurs when $\mu > 0$).

\section{Dilepton signal from chargino pair production}

We see from Fig.~1 that at least for the large values of $\mu$ expected
in SUGRA models, along with $\sigma(p\bar p\to\tw_1\tz_2 X)$,
the $p\bar p\to\tw_1\overline{\tw_1}X$ cross section remains
large out to large values of $m_{\tg}$. A signal (with the expected
strength) in this channel
accompanying the clean trilepton events discussed in Section II would
serve as strong evidence for the supersymmetric origin of these events.
The best signature for chargino
pair production is in the clean, opposite-sign (OS), isolated
dilepton channel\cite{EARLY,ANTRI,NEW,LOPEZ};
single lepton plus jets and multi-jet $+\eslt$ signatures should be well
below SM backgrounds. Assuming that it is possible to veto $Z \to
\ell\bar{\ell}$ events,
the major SM backgrounds to the clean OS dilepton
signal consist of Drell-Yan $\ell\bar{\ell}$ production,
$\gamma^*,Z\to\tau\bar\tau$ production, $t\bar t$ production (where the
$b$ jets are soft), and
pair production of vector bosons, especially $WW$ production.

Since clean dilepton events can come from a variety of SUSY sources,
to evaluate the signal we generate {\it all} possible SUSY production
reactions.
We implement the following set of cuts, designed to extract signal
from these backgrounds.
The results of these successive cuts on two signal cases, and backgrounds,
are shown in Table II.
\begin{itemize}
\item We require exactly two {\it isolated} OS (either $e$ or $\mu$ ) leptons
in each event, with $p_T(\ell_1)>10$ GeV and $p_T(\ell_2)>7$ GeV, and
$|\eta (\ell ) |<2.5$. In addition, we require {\it no} jets, which
effectively reduces most of the $t\bar t$ background. These cuts are
labelled as ($2\ell$, $0j$).
\item We require $\eslt >25$ GeV (labelled ($\eslt$)). This cut removes
backgrounds from Drell-Yan dilepton production, and also
the bulk of the background from $\gamma^*, Z\to\tau\bar{\tau}$ decay. Notice
that the $\gamma^*,Z\to\tau\bar{\tau}$ background still dominates.
\item We require $\phi (\ell\bar{\ell})<150^0$, to further reduce
$\gamma^*,Z\to\tau\bar{\tau}$ (labelled $\phi$).
\item We require the $Z$ mass cut:
invariant mass of any opposite-sign, same flavor dilepton pair not reconstruct
the $Z$ mass, {\it i.e.} $m(\ell\bar{\ell})\ne M_Z\pm 10$ GeV. This cut
is labelled ($M_Z$).
\end{itemize}
We see from the fourth row of Table II
that the dominant remaining background comes from
$WW$ production. However, the leptons and $\eslt$ from two-body $W$ boson
decays are considerably harder than those from three body chargino decay to
a massive LSP, via $\tw_1\to e\nu_e\tz_1$. It has been shown in Ref. \cite{BST}
that a cut on $B=|\vec{\eslt}|+|p_T(\ell_1)|+|p_T(\ell_2)|<100$ GeV was
effective to separate softer top {\it squark} events from hard top {\it quark}
events; the same works here to gain significant rejection on $WW$ BG
with only modest loss of signal, as shown in Fig. 6.
(If the decay channel $\tw_1\to\tz_1 W$ is open, this
cut is not effective; we have, however, checked that these cases are very
difficult to see above background anyway.) Hence, we further require
\begin{itemize}
\item $B=|\vec{\eslt}|+|p_T(\ell_1)|+|p_T(\ell_2)|<100$ GeV,
(labelled as ($B$)).
\end{itemize}
The final cross sections for two cases of signal and background are
shown in the last row of Table II.
Assuming an integrated luminosity of 25 $fb^{-1}$, both
these signal cases should be observable above background
at the $5\sigma$ level.
For the second case, this corresponds to 214
signal events, as opposed to $\sim$1100 BG events- a 20\% effect.
While this may sound marginal,
we note that the bulk of the background, from $WW$ production, can be
well-measured and normalized, so that the signal can be looked for as a
distortion in, for instance, the low energy end of the distribution in $B$
shown in Fig.~6.

We plot the regions of observability of this signal
in the $m_0\ vs.\ m_{1/2}$ SUGRA plane, in the upper frames {\it a})
of Figs.~7-9 for the same values of parameters as in Figs.~3-5. The hatched
and bricked regions are identical to those in Figs.~3-5.
The following points are worth noting.
\begin{itemize}
\item For the $\tan\beta=2$, $\mu <0$ case illustrated in Fig. 7{\it a},
the region of space explorable via clean dileptons is a subset of the region
accessible via clean trileptons. However, for the $\tan\beta=2$, $\mu >0$
case shown in Fig 8{\it a}, the clean dilepton channel offers a reach well
into the ``hole'' region of the trilepton plot Fig. 4{\it a}.

\item There are significant regions of the $m_0\ vs.\ m_{1/2}$ plane where
both the trilepton as well as the clean dilepton signal should be observable.
As noted above, the simultaneous detection of a signal in both channels
at the expected relative rates, and with the correct kinematics, may serve to
identify their origin. It is also instructive to note that charginos should
also be detectable at LEP II over much of this same region. Although this will
obviously be sensitive to the energy at which LEP II becomes operational,
it seems likely that the reach of MI and \tevst in the clean dilepton
channel will not significantly exceed that of LEP II.
\end{itemize}

\section{Confirmation in OS dilepton plus jets channel}

If a signal is detected in the $\tw_1\tz_2\to 3\ell +\eslt$, then
as a confirmation, there ought to be as well a signal in the
$\tw_1\tz_2\to q\bar{q'}\tz_1 +\ell\bar{\ell}\tz_1$ channel.
For this
reaction, one ought to detect an OS but same flavor dilepton pair with
invariant mass bounded by $m_{\tz_2}-m_{\tz_1}$, recoiling against
one or two jets, plus moderate $\eslt$. Although the signal rate is expected
to
exceed that in the trilepton channel, substantial backgrounds
from $\gamma^*,Z\to\tau\bar{\tau}$, $t\bar t$ and vector boson pair production
make detection in this channel more difficult.

Nonetheless, it is interesting
to identify regions of SUGRA parameter space where
the dilepton plus jets channel yields a confirmatory signature.
Again, we must generate {\it all} possible SUSY reactions.
The following set of cuts are implemented, and the results tabulated in
Table III.
\begin{itemize}
\item We require two {\it isolated} OS ({\it same} flavor) dileptons, plus
at least one detectable jet.
As before, the leptons are required to satisfy $p_T(\ell_1)>10$ GeV and
$p_T(\ell_2)>7$ GeV, and
$|\eta (\ell)|<2.5$.
These cuts are labelled as ($2\ell$, $nj$).
\item Requiring $\eslt >25$ GeV (labelled ($\eslt$)) removes
much of the background from $\gamma^*,Z\to\tau\bar{\tau}$ decay and
also SM Drell-Yan dilepton production.
\item We require $m(\ell\bar{\ell})<80$ GeV, to remove backgrounds from
$Z$ decays, and much of $t\bar{t}$ production.
This cut is labelled ($m_{\ell\bar{\ell}}$).
\item We require $\phi (\ell\bar{\ell})<90^0$, to further reduce
$Z\to\tau\bar{\tau}$ (labelled $\phi$).
\item Finally, to remove much of the remaining $t\bar t$ background, we
require $p_T(fast\ jet)<50$ GeV (labelled $p_T^j$), and invariant mass
of all reconstructed jets to be $m(jets)<70$ GeV (labelled $m_{jets}$).
\end{itemize}
The residual backgrounds after these
cuts mainly come from vector boson pair production and from
$\tau\bar{\tau}$ pair production, and
sum to 25.6~$fb$. For 25~$fb^{-1}$ of integrated
luminosity, the signal cross section required for
a $5\sigma$ observation is 5.1~$fb$. Both of the signal
cases listed in Table III are thus observable. We remark again that
the main background processes will be well measured from collider data.
In addition, since the signal only occurs in the same-flavor channel,
it leads to an excess of $e\bar e$ and $\mu\bar{\mu}$ events over
$e\bar{\mu}+\bar{e}\mu$ events.

The regions of the SUGRA plane where this signal is observable
are shown in Fig.~7{\it b}-9{\it b} for the same
values of parameters as in Figs.~7{\it a}-9{\it a}.
We see that except for tiny regions of parameter space,
this signal does not probe regions not accessible via the other channels.
There are, however, sizeable regions where the jets plus OS dilepton signal
can be used to confirm a signal seen in other channels.

Finally, we remark
that a simultaneous measurement of the SUSY cross section in this channel
together with $\sigma(3\ell)$ discussed in Sec. II, directly yields
the leptonic branching fraction for decays of the chargino. This assumes
that the sample of jets plus OS dilepton events comes mainly from
$\tw_1\tz_2$ production; {\it i.e.} contributions from other SUSY sources are
negligible, and further, that the background can be reliably subtracted.
It would thus be interesting to see if cuts can be devised to separate
the $\tw_1\tz_2$ source of these dilepton
events from cascade decays of squarks and gluinos
or from associated production processes as was done for the trilepton
events in Ref.\cite{BCPT}

\section{Other Search Strategies and Concluding Remarks}

\subsection{Summary of signals from charginos and neutralinos}

In supersymmetric models where the running gaugino masses unify at a high
scale, and the weak scale parameter $\mu$ greatly exceeds the
$U(1)$ and $SU(2)$ gaugino masses $M_1$ and $M_2$ (as is the case
for the minimal SUGRA model), one typically
expects $m_{\tw_1} \sim m_{\tz_2}\sim ({1\over 3}-{1\over 4})m_{\tg}$.
As a result, $p\bar{p} \to \tw_1\tz_2$ and $p\bar{p} \to \tw_1\overline{\tw_1}$
are the dominant sparticle production processes at the Tevatron if
gluinos and squarks are heavier than 250-300~GeV, so that signals from these
channels offer the best reach for supersymmetry at high luminosity upgrades
of the Tevatron. The most
promising signals from these reactions are:
\begin{itemize}
\item clean ({\it i.e.} jet-free), isolated trilepton events
from $\tw_1\tz_2\to 3\ell +\eslt$,
\item clean, OS dilepton events from
$\tw_1\overline{\tw_1}\to \ell\ell' +\eslt$, and
\item OS but same flavour dilepton plus jets events from
$\tw_1\tz_2\to q\bar{q'}\ell\bar{\ell}+\eslt$.
\end{itemize}

Motivated by the recent interest in the \tevst upgrade of
the Tevatron collider--- designed to yield an integrated luminosity of $\sim 25
fb^{-1}$--- we have used ISAJET to compute these SUSY
signals within the framework
of the minimal supergravity model with gauge coupling unification and
radiative breaking of electroweak symmetry.
Specifically, we have scanned sections of the multi-dimensional
parameter space, and delineated the regions where the SUSY signal is
observable over SM {\it physics} backgrounds.
Since we have not included
a real detector simulation, effects due to multiple scattering during
bunch crossings, or backgrounds due to $\pi$'s or $\gamma$'s faking
an electron (these are all sensitive to details of the ultimate collider and
detector design), our conclusions must be regarded as on the optimistic side.

The region of the $m_0-m_{1/2}$ supergravity parameter plane where
the trilepton signal is expected to be
observable, with a significance $\geq 5\sigma$ at the Main Injector
and at the \tevst, is shown by solid squares and hollow squares, respectively,
in Figs.~3-5{\it a}. Figs.~7-9 illustrate the corresponding regions where
the dilepton signals are expected to be observable.
{}From these figures, we see that
experiments at \tevst should be able to probe significantly larger
regions of parameter space than at the Main Injector, in both the dilepton
and trilepton channels. When $m_0$ is small ($\leq 200$~GeV),
the trilepton signal may be (indirectly) sensitive to gluino masses
as heavy as $\sim 700$~GeV! However, even in this favorable region, there exist
holes of non-observability due to neutralino/chargino
decays to soft or invisible particles. For $\tan\beta =2$, $m_0$
large ($> 400$~GeV) and negative values of $\mu$, experiments at
\tevst may be able to see trilepton signals essentially until
the ``spoiler mode''
$\tz_2\to\tz_1 H_{\ell}$ becomes accessible
($m_{\tg}\sim 500$ GeV). However, if $\mu >0$, large interference
effects in the neutralino leptonic decay width severely suppress the signal
if 400~GeV~$\alt m_0 \alt$~1000~GeV,
and a gaping hole where there is no detectable trilepton
signal remains in the parameter space, all the way down to
the currently excluded LEP chargino mass limit. For large values of
$\tan\beta$, the parameter space hole is somewhat smaller, though generally
speaking the cross section often falls below the detectable level even when
the spoiler decay modes of $\tz_2$ are closed.

If a trilepton signal is discovered, will there be a confirming discovery in
the dilepton channels? Are clean
dilepton signals observable in the parameter
space hole mentioned above?
Figs. 7-9 show that indeed large regions of parameter
space can be probed
in the dilepton channel. The region probed by the OS dilepton plus
jets channel forms a subset of that explorable via trileptons.
However, the clean (no jets) OS dilepton channel is to some extent
complementary to the trilepton channel, in that this signal
can be seen in regions
of parameter space where the trilepton signal is suppressed. We note,
however, that the regions explorable via dileptons at \tevst overlap
considerably with the regions explorable by LEP II via slepton and
chargino searches, as shown by the dashed contours in Figs. 3{\it b}-5{\it b}.

\subsection{Comparison with signals from other channels}

How does the reach for supersymmetry in the OS dilepton and trilepton channels
considered here compare with the reach in other channels such at
multi-jets $+\eslt$ and same-sign dileptons?
The various multi-jet and multi-lepton signals from production and cascade
decay of all SUSY particles have been computed previously in Ref.~\cite{BKT}
for $\mu =-m_{\tg}$. We have repeated this calculation for
comparison with the results obtained here, except now we use the SUGRA
parameters with $m_0=m_{1/2}$ (which yields $m_{\tq}\sim m_{\tg}$) and
$m_0=4m_{1/2}$ (which yields $m_{\tq}\sim (1.5-1.7) m_{\tg}$, and consider
both signs of $\mu$. We set $A_0=0$, and $\tan\beta=2$.
We have computed the cross sections
for the multi-jet $+\eslt$ and SS dilepton event topologies, with the
same cuts as in Ref.\cite{BKT}
for various values of $m_{1/2}$ ($\sim {1\over 3}m_{\tg}$) and
compared against SM backgrounds\cite{BKT} for $m_t=170$ GeV.
The reach of the Tevatron MI,
and also \tevst, computed in terms of $m_{\tg}$, is shown in
Table IV, for comparison with results obtained in this paper. The reader
should view these numbers in proper perspective, since we have not scanned
the complete parameter space in constructing this Table, but have fixed
$A_0$ and $\tan\beta$. Indeed, the reach in some channels, {\it e.g.} the
trilepton channel, is sensitive especially to $\tan\beta$.

For the multi-jet $+\eslt$ channel (labelled $\eslt$), naive
application of the $5\sigma$ criterion gives a reach of typically 350-400~GeV
at the TeV$^*$ (listed in parenthesis). However, since the signal to background
ratio $\frac{S}{B}$ is smaller than 3\%
(if the signal is at the $5\sigma$ level),
and there are no characteristic kinematic bumps expected in the signal,
it may not be realistic to expect detectability at this level. We, therefore,
also list a conservative reach by requiring, in addition, ${S\over B} >0.25$
(no parenthesis).

The most peculiar point about the $\eslt$ signal is that in the trilepton
``hole'' region, for the MI at least, there is again no reach
beyond the current LEP chargino bound. This is because for $\mu > 0$ and
small values of $\tan\beta$, the
LEP chargino limit already implies
$m_{\tg}\agt 250$ GeV, for which the strong production
cross sections are already rather small\cite{FN3}.
As can be seen from Fig. 2{\it b},
chargino/neutralino pair production is the dominant production
mechanism, and most multi-jet $+\eslt$ events come from their hadronic
decays. Since $\tw_1$ and $\tz_{1,2}$ tend to be especially light
for positive values of $\mu$,
their production yields a rather soft
$\eslt$ spectrum, which is difficult to see above background.

We also list the reach for supersymmetry in the SS dilepton channel. This
signal was originally proposed as a way to test the Majorana nature of the
gluino, by searching for $\tg\to q\bar{q'}\tw_1$ decays, followed by
$\tw_1\to\ell\nu\tz_1$. The $\tg\tg\to\ell^\pm\ell^\pm +jets+\eslt$ channel is
of course only useful as long as one is making a sufficient number of
gluino pairs in the first place, and that there is a reasonable efficiency
to detect the decay leptons. We see from Table IV that the SS dilepton
signal is as well not visible in the trilepton ``hole'' region, due to a
combination of low production cross section together with
a low efficiency to detect the leptons which tend to be soft because
both $m_{\tw_1}$ and $m_{\tw_1}-m_{\tz_1}$ tend to be smaller for $\mu > 0$
as compared to for $\mu < 0$.
Indeed, for the
other parameters of Table IV where there is a reach via SS dileptons up to
quite high values of $m_{\tg}$, most of the dileptons originate
from $3\ell$ events, where one of the leptons is lost.
We further see from Table IV that the reach in SS or OS dileptons is
generally similar, and that in moving from Tevatron MI to \tevst, the
reach in $m_{\tg}$ is increased by typically 100~GeV.

Slepton
pair production (which has been included in our computation)
is best detected via the clean OS lepton signal\cite{SLEP}. The slepton
signal will be observable at the MI only for slepton masses lighter
than around 50~GeV\cite{SLEP}, while the reach of \tevst is
comparable to that of LEP II.

A comparison of the
reach of every channel shows that the maximal reach is always obtained in the
clean $3\ell$ channel, except for the ``hole'' region where
$m_0\agt 400$~GeV,
and $\mu >0$: in this region, only \tevst has a significant reach
in the clean OS dilepton channel, and perhaps, in the $\eslt$ channel.

\subsection{Conclusions}

The current run of the Tevatron $p\bar p$ collider, run IB, is expected
to attain $0.1\ fb^{-1}$ of integrated luminosity per experiment. With such
a data sample, the CDF and D0 experiments should be sensitive
to SUSY signals in several channels
other than the canonical multi-jet $+\eslt$ channel.
In particular, the clean trilepton channel ought to allow Tevatron experiments
to probe regions of parameter space significantly beyond the range of LEP,
and perhaps, even up to the reach of LEP II.

In 1996, LEP II is expected to ramp up to CM energy $\sqrt{s}\sim 175-190$~GeV,
which should allow $m_{\tell}$, $m_{\tw_1}$ and $m_{H_{\ell}}\sim 80-90$~GeV
to be probed. The relatively model-independent search for supersymmetry
will optimistically probe the regions below the dashed contours in
Figs. 3{\it b}-5{\it b}. By 1999, the Tevatron MI may be operational, and
is expected to amass $\sim 1\ fb^{-1}$ of data per experiment. The regions
explorable at the MI can be significantly larger than
those accessible to LEP II,
but only in the lower regions of $m_0$, where sleptons are significantly
lighter than squarks, so that leptonic decays of neutralinos are enhanced.
The reach of the Tevatron MI via dilepton signals
is essentially a subset of the
parameter space that LEP II can explore, so we do not expect new progress
to be made in this channel by MI experiments,
above and beyond what LEP II can do.

An upgrade of the Tevatron to \tevst would be particularly well-suited to
exploit the clean trilepton channel expected from many SUSY models.
The relatively small signal rates may well be detectable above tiny SM
backgrounds, if machine and detector dependent backgrounds can be controlled.
A luminosity upgrade to $\sim 25\ fb^{-1}$ of integrated luminosity ought to
allow for a substantial increase in the amount of SUSY parameter space that can
be explored: in favourable
cases, equivalent gluino masses as high as 700 GeV can be
probed! However, in spite of its increased reach for SUSY,
we note that there are substantial regions of SUSY
parameter space which {\it cannot} be explored by \tevst, either due
to invisible neutralino decays, leptonic branching fraction suppression,
the turn on of neutralino spoiler modes, or just kinematically suppressed
production cross sections.

Probably the main motivation for weak scale supersymmetry comes from the
fact that it provides a mechanism for stabilizing the weak scale if sparticle
masses are smaller than $\sim 1$~TeV. Some authors have attempted
to quantify this by imposing fine-tuning requirements, and have obtained
upper bounds on sparticle masses\cite{DIEGO}. Their arguments,
although admittedly rather subjective, nonetheless suggest sparticle mass
bounds $m_{\tg},m_{\tq}\alt 700-800$ GeV.
Even taking such an upper limit
literally, it would still not be possible to make a definitive seach for
supersymmetry at LEP II or at \tevst. On the other hand, recent
work\cite{ATLAS,BCPT2} on the multi-jet $+\eslt$ signal indicates
that the CERN LHC
$pp$ collider, operating at $\sqrt{s}=14$ TeV, can explore the entire range
of parameter space up to $m_{\tg}\sim 1300$ GeV ($m_{\tq}>>m_{\tg}$) or
$m_{\tg}\sim 2000$ GeV ($m_{\tq}\sim m_{\tg}$). It thus appears that
experiments at supercolliders are essential to ensure that weak scale SUSY
does not escape experimental detection.

\acknowledgments

We thank Kaushik De, Teruki Kamon, and the Fermilab \tevst working group for
motivating this project.
We thank A. Bartl, and also M. Brhlik,
for independently confirming our branching fraction calculations.
This research was supported in part by the U.~S. DOE
grants DE-FG-05-87ER40319, DE-FG-03-94ER40833, and DE-FG-02-91ER40685.
%

%
\newpage

\begin{table}
\caption[]{Cross section in $fb$ for the clean trilepton signal
before and after successive cuts for
two signal cases, and major physics backgrounds. The signal cases list
the value of ($m_0, m_{1/2}, sgn(\mu $)), and take $A_0=0$ and $\tan\beta =2$.
}

\bigskip

\begin{tabular}{cccccc}
cuts & ($200,165,-1$) & ($200,200,+1$) & $t\bar t(170)$ & $WZ$ & $total\ BG$ \\
\tableline
$3\ell$ & 6.3 & 4.3 & 0.29 & 26.9 & 27.2 \\
$3\ell ,\eslt$ & 5.3 & 4.3 & 0.26 & 21.1 & 21.4\\
$3\ell ,\eslt ,M_Z$ & 5.0 & 3.4 & 0.2 & 0.35 & 0.55 \\
$3\ell ,\eslt ,M_Z, 0j$ & 2.4 & 1.7 & 0.005 & 0.2 & 0.205 \\
\end{tabular}
\end{table}

\begin{table}
\caption[]{Cross section in $fb$ for the clean, OS dilepton signal,
before and after successive cuts for
two signal cases, and major physics backgrounds. The signal cases list
the value of ($m_0, m_{1/2}, sgn(\mu $)), and take $A_0=0$ and $\tan\beta =2$.
}

\bigskip

\begin{tabular}{ccccccccc}
cuts & ($200,100,+1$) & ($600,80,+1$) & $t\bar t(170)$ &
$\gamma^*,Z\to\tau\bar{\tau}$ & $WW$ & $WZ$ & $ZZ$ & $total\ BG$ \\
\tableline
$2\ell ,0j$ & 130.7 & 57.5 & 0.24 & 10203 & 211.4 & 5.1 & 12.6 & 10432.3 \\
$2\ell ,0j,\eslt$ & 41.2 & 13.6 & 0.2 & 444.1 & 150.2 & 3.2 & 9.5 & 607.2 \\
$2\ell ,0j,\eslt ,\phi$ & 34.0 & 9.5 & 0.15 & 4.7 & 126.1 & 2.4 & 7.1 & 140.4
\\
$2\ell ,0j,\eslt ,\phi ,M_Z$ & 33.9 & 9.3 & 0.1 & 4.7 & 118.8 & 0.7 & 0.1 &
124.4 \\
$2\ell ,0j,\eslt ,\phi ,M_Z, B$ & 30.1 & 8.6 & 0.02 & 4.7 & 38.7 & 0.3
& 0.1 & 43.8 \\
\end{tabular}
\end{table}
\begin{table}
\caption[]{Cross section in $fb$ for the OS same-flavor dilepton plus
jets signal,
before and after successive cuts for
two signal cases, and major physics backgrounds. The signal cases list
the value of ($m_0, m_{1/2}, sgn(\mu $)), and take $A_0=0$ and $\tan\beta =2$.
The BG entry VV sums over $WW$, $WZ$ and $ZZ$ production.
}

\bigskip

\begin{tabular}{ccccccc}
cuts & ($200,150,-1$) & ($300,100,+1$) & $t\bar t(170)$ &
$\gamma^*,Z\to\tau\bar{\tau}$ & $VV$ & $total\ BG$ \\
\tableline
$2\ell ,nj$ & 32.2 & 28.3 & 134 & 893 & 264.4 & 1291.4 \\
$2\ell ,nj,\eslt$ & 25.6 & 12.5 & 119.6 & 281 & 93.1 & 493.7 \\
$2\ell ,nj,\eslt ,m_{\ell\bar{\ell}}$ & 25.5 & 12.5 & 58.9 & 281 & 46.9 &
386.8 \\
$2\ell ,nj,\eslt ,m_{\ell\bar{\ell}},\phi$ & 14.7 & 8.1 & 36 & 48.6 &
22.3 & 106.9 \\
$2\ell ,nj,\eslt ,m_{\ell\bar{\ell}},\phi ,p_T^j$ & 7.8 & 5.9 & 4.6 &
13.1 & 16.9 & 34.6 \\
$2\ell ,nj,\eslt ,m_{\ell\bar{\ell}},\phi ,p_T^j,m_{jets}$ & 6.2 & 5.2 &
1.7 & 9.0 & 14.9 & 25.6 \\
\end{tabular}
\end{table}
\begin{table}
\caption[]{Approximate reach in terms of $m_{\tg}$ (GeV) via
various event topologies
for {\it a}) Tevatron Main Injector and
{\it b}) \tevst, assuming $A_0=0$ and $\tan\beta =2$, with other SUGRA
parameters as listed.
We use $m_t =170$ GeV for the background.
None means that there is no reach
beyond the current LEP bound.}

\bigskip

\begin{tabular}{cccccc}
case & $E\llap/_T$ & $OS\ (0j)$ & $OS\ (nj)$ & $SS$ & $3\ell$ \\
\tableline
$a)\ {\rm MI}\ \ (1fb^{-1})$ & & & & & \\
$m_0\sim m_{1/2}\ (\mu <0)$ & 310 (330) & 275 & 330 & 330 & 450 \\
$m_0\sim m_{1/2}\ (\mu >0)$ & 320 (345) & 400 & 360 & 370 & 600 \\
$m_0\sim 4m_{1/2}\ (\mu <0)$ & 230 (260) & 180 & 270 & 270 & 310 \\
$m_0\sim 4m_{1/2}\ (\mu >0)$ & none (none) & none & none & none & none \\
\tableline
$b)\ {\rm TeV}^*\ \ (25fb^{-1})$ & & & & & \\
$m_0\sim m_{1/2}\ (\mu <0)$ & 310 (400) & 380 & 440 & 430 & 700 \\
$m_0\sim m_{1/2}\ (\mu >0)$ & 320 (430) & 500 & 490 & 480 & 700 \\
$m_0\sim 4m_{1/2}\ (\mu <0)$ & 230 (345) & 300 & 360 & 370 & 500 \\
$m_0\sim 4m_{1/2}\ (\mu >0)$ & none (365) & 300 & none & none & none \\
\end{tabular}
\end{table}

%
\newpage
\begin{figure}
\caption[]{Cross sections at the Tevatron ($\sqrt{s}=2$ TeV)
for total $\tg\tg +\tg\tq +\tq\tq$ production, associated production,
and $\tw_1\tz_2$ as well as $\tw_1\overline{\tw_1}$ and $\tw_1\tz_1$
production,
as a function of $m_{\tg}$.
In {\it a}), we take $m_{\tq}=m_{\tg}$, while in {\it b}) we take
$m_{\tq}=2 m_{\tg}$. We have also taken $\tan\beta =2$ and $\mu =-m_{\tg}$.
}
\end{figure}
%
\begin{figure}
\caption[]{Same as Fig. 1, except $\mu =+m_{\tg}$.
}
\end{figure}
%
\begin{figure}
\caption[]{In {\it a}), we show regions of the $m_0\ vs.\ m_{1/2}$
plane where supersymmetry should be detectable at the Tevatron MI
(black squares), and at \tevst at $10\sigma$ (squares with x) and
$5\sigma$ levels (empty squares), by searching
for clean trilepton events. The bricked region is excluded by theoretical
constraints, while the gray shaded region is excluded by experiment.
We take $\tan\beta =2$, $A_0=0$, $\mu <0$, and $m_t =170$ GeV.
In {\it b}), we show various contours for $m_{\tg}$, $m_{\tw_1}$, and
$m_{\tell_R}$ for comparison with the results of {\it a}).
We also show via the dashed contours in {\it b}) the approximate reach of
LEP II.
}
\end{figure}
%
\begin{figure}
\caption[]{Same as Fig. 3, except now $\mu >0$.}
\end{figure}
%
\begin{figure}
\caption[]{Same as Fig. 3, except that $\tan\beta =10$.}
\end{figure}
%
\begin{figure}
\caption[]{Distribution in $B=|\vec{\eslt}|+|p_T(\ell_1)|+|p_T(\ell_2)|$,
for $WW$ background (histogram), and the two signal cases of Table 2,
after the first four cuts of Table 2.}
\end{figure}
%
\begin{figure}
\caption[]{In {\it a}), we show regions of the $m_0\ vs.\ m_{1/2}$
plane where signals should be detectable at the Tevatron MI
(black squares), and at \tevst at $10\sigma$ (squares with x) and
$5\sigma$ levels (empty squares), by searching
for clean {\it dilepton} events. The bricked region is excluded by theoretical
constraints, while the gray shaded region is excluded by experiment.
We take $\tan\beta =2$, $A_0=0$, $\mu <0$, and $m_t =170$ GeV.
In {\it b}), we show similar regions for the {\it dilepton plus jets}
signal.
}
\end{figure}
%
\begin{figure}
\caption[]{Same as Fig. 7, except $\mu >0$.}
\end{figure}
%
\begin{figure}
\caption[]{Same as Fig. 7, except $\tan\beta =10$.}
\end{figure}


\begin{references}
\bibitem{FN1} Other reactions, such as $\tz_1\tz_2$ and $\tz_2\tz_2$
production, occur
at much smaller rates when $|\mu |$ is large, as in minimal SUGRA models.
\bibitem{FN2} In addition there may be an observable rate for clean
like-sign dilepton events from $\tw_1\tz_2$ when one of the three leptons
escapes detection. This signal cannot, however,
be used as a confirmatory channel since it has exactly the same origin as the
trilepton signal. It would not be unreasonable to add this cross section
to the trilepton cross section in evaluating the reach of the \tevst since
both these channels are free of physics backgrounds. We have conservatively
not done so in our analysis.
\bibitem{FN} The result presented here is at variance with
results presented in Ref. \cite{LOPEZ,KAMON,KANE},
where the authors suggest that
trilepton searches will significantly extend the reach in $\tw_1$
independently of the sign of $\mu$.
\bibitem{FN3} Of course, Tevatron experiments should independently search
for gluinos, regardless of the LEP constraints on the chargino mass since
the gaugino mass unification assumption could turn out to be false.
\end{references}
\end{document}